# Efficient and Reliable Topology Control based Opportunistic Routing Algorithm for WSNs

Ning Li, *Student Member, IEEE*, Jose-Fernan Martinez-Ortega, Vicente Hernandez Diaz

*Abstract*—The opportunistic routing has advantages on improving the packet delivery ratio between source node and candidate set (PDRsc). However, considering the frequent topology variation in wireless sensor networks, how to improve and control the PDR has not been investigated in detail. Therefore, in this paper, we propose an efficient and reliable topology control based opportunistic routing algorithm (ERTO) which takes PDRsc into account. In ERTO, the interference and transmission power loss are taken into account during the calculation of PDRsc. The PDRsc, the expected energy consumption, and the relationship between transmission power and node degree are considered to calculate the optimal transmission power and relay node degree jointly. For improving the routing effective and reducing the calculation complexity, we introduce the multi-objective optimization into the topology control. During the routing process, nodes calculate the optimal transmission power and relay node degree according to the properties of Pareto optimal solution set, by which the optimal solutions can be selected. Based on these innovations, the energy consumption, the transmission delay, and the throughout have been improved greatly compared with the traditional power control based opportunistic routing algorithms.

*Index Terms*—Topology control, opportunistic routing, wireless sensor network, multi-objective optimization.

## I. INTRODUCTION

In recent years, the wireless sensor networks (WSNs) have been used widely, such as in battlefield [1][2], in vehicle networks [3][4][5], in underwater cooperation robot networks [6][7][8], etc. In WSNs, one of the important issues is the routing protocol design, which guarantees reliable and efficient data transmission from source node to destination node. There are two different kinds of routing strategies: deterministic routing and opportunistic routing. The deterministic routing algorithms select one optimized route before transmission [9]; the main disadvantage of deterministic routing is that it simply applies the operations and principles inherited from legacy routing solutions which were initially conceived for wired networks, so it cannot adapt well on the dynamic environment variation. Thus, the opportunistic routing has been proposed. The opportunistic routing regards the shared wireless medium as an opportunity rather than a limitation for data packet transmission [10]. The opportunistic routing overcomes the drawbacks of deterministic routing by taking advantage of the broadcast nature of wireless medium.

### A. Motivation

The opportunistic routing can improve the network capability successfully compared with the deterministic routing, especially the packet delivery ratio between the sender and the candidate set (PDRsc). The PDRsc is defined as the probability that the data packet sent by sender can be received by at least one relay node in candidate set. In previous works, such as [11], [12], [13], [14], [15], [16], [17], [18], the PDRsc has been utilized in routing algorithm design. However, in the WSNs, the PDRsc changes when the network topology changes; how to control and improve the PDRsc and how the PDRsc affects the routing performance have not been investigated detailed. There are two parameters can affect the performance of PDRsc in opportunistic routing (the number of node in candidate set and the packet delivery ratio between sender and one of its neighbors (PDRsn)) [12][18], so for improving the routing performance, the opportunistic routing algorithm should be able to aware and control: 1) *number of relay nodes*: in opportunistic routing, large number of relay nodes means high PDRsc; however, the energy consumption and interference increase when the number of relay nodes is large; so the number of relay nodes in opportunistic routing should in an appropriate level; 2) *link availability*: whether the transmitted data packet can be received by the receiver successfully relates to both the transmission power loss and the interference of the receiver [19]; so for improving the PDRsc, the algorithm should be able to adaptive the changing of transmission power loss and interference, especially in WSNs.

In this paper, we use the topology control to achieve the objectives introduced above. The topology control has been applied into opportunistic routing and many high quality algorithms have been proposed. In these algorithms, the transmission power can be adjusted based on the network condition during the routing process. During the topology control, on one hand, the more network parameters are taken into account, the more accurate of the topology control is. However, when taking too many parameters into account, there will be: 1) the calculation of the topology control becomes too complexity to be applied in practice; 2) getting the optimal solution for all these network parameters is not always feasible, such as the minimum energy consumption and the minimum transmission area. On the other hand, the traditional topology control algorithms calculate the optimal transmission power for each node based on the optimal algorithms, and once the current transmission power does not equal to the optimal one, the nodes change their transmission power. This approach is effective when the network resources, such as the bandwidth and the link capability, are abundant; however, in wireless sensor network, where the network resources are limited and the network topology changes frequently due to node mobility or failure, the heavy control cost will consume a plenty of network resource which could have been used in data packets transmission. So the topology control algorithm should be able to reduce the extra control cost as far as possible. In this paper, we introduce the multi-objective optimization into the topology control. This algorithm can not only find the optimal tradeoff between these network parameters, but also reduce the control cost.

During the topology control, the relationship between transmission power and the number of nodes in candidate set should be paid attention. Considering the conclusions in [21], the

Ning Li, Jose-Fernan Martinez-Ortega, and Vicente Hernandez Diaz are with the Universidad Politecnica de Madrid, ETSI y Sistemas de Telecomunicacion, Campus Sur, Ctra. de Valencia Km 7, 28031, Madrid, Spain. Email: {li.ning; jf.martinez; vicente.hernandez}@upm.es.

number of nodes in the coverage area under specific transmission power follows a Passion distribution when the nodes are uniformly distributed. So during the topology control, the transmission power and the number of nodes in candidate set should be optimized jointly rather than separately. For instance, if the optimal transmission power is *P* and the optimal relay node number is *n* by mathematic calculation, however, the combination of this transmission power and number of nodes in candidate set may not exist or exists with low probability according to the conclusion in [21]; then the routing performance will deteriorate. In this paper, the number of relaying nodes and the transmission power are optimized jointly.

*B. Main Contributions*

In this paper, we focus on finding an efficient and reliable topology control based opportunistic routing (ERTO) for wireless sensor network, which can improve the successful data packet delivery ratio while reducing the energy consumption. The main contributions of this paper are as follows:

1. In this paper, we propose an accurate calculation model for the PDRsc, which takes the network interference into consideration; moreover, we also define the relay node degree and candidate relay area for opportunistic routing and propose the calculation model for candidate relay area;
2. Based on the physical relationship between transmission power and relay node degree, we optimize these two parameters jointly; moreover, taking the PDRsc, the expected energy consumption, and the relation between transmission power and relay node degree into account, for getting an optimal tradeoff between these parameters, we introduce the multi-objective optimization approach into the topology control; by this approach, more parameters can be taken into account during the topology control than the traditional approach without increasing the complexity of the algorithm; moreover, an optimal tradeoff between these parameters can be calculated;
3. We propose an efficient and reliable topology control based opportunistic routing algorithm for wireless sensor networks. In ERTO algorithm, the transmission power of node is adjusted only when the current transmission power and relay node degree are not in Pareto optimal solution set; the optimal transmission power and relay node degree are determined based on the characteristic of Pareto optimal solution set. By this approach, the control cost has been reduced.

The rest of the paper is organized as follows: in Section II, the related works are introduced briefly; Section III introduces the network model used in this paper; Section IV introduces the optimization model and analyzes the optimization model in detail; in Section V, based on the conclusions of Section IV, an efficient and reliable topology control based opportunistic routing algorithm is proposed; Section VI evaluates the performance of ERTO and compares it with ExOR, EEOR, TCOR; Section VII concludes our works in this paper.

## II. RELATED WORKS

There are many opportunistic routing algorithms have been proposed, such as [10, 22-27]. In [10], the S. Biswas and R. Morris first proposed the concept of opportunistic routing, named ExOR. In ExOR, the routing protocol and the MAC layer operations are integrated together. The source node broadcasts a batch of packets to a list of nodes which can potentially forward these packets. Each neighbor in the forwarding list waits for its turn to transmit the received packets by using the same transmission strategy as source node. Bletsas et al. proposed ILOR in [18]. The ILOR is an opportunistic relaying protocol for noise and interference limited slow fading environments. In ILOR, the relay selection is based on the estimation of link quality towards to the destination. Each relay node estimates its own RSSI by listening to a pilot signal from destination [19]. The Simple and Practical opportunistic routing (SPOR) algorithm has been proposed in [20]. The SPOR is designed for multi-hop wireless networks. In SPOR, the node iteratively forwards data packet and acknowledges its reception at each hop. In [21], the authors proposed the Parallel-OR, which supports multiple simultaneous flows in large wireless networks. The Parallel-OR is massively parallel since it is performed by many nodes simultaneously to maximize the opportunistic gain. In OPSR [22], the opportunistic forwarding and packet scheduling are combined to support multiple simultaneous flows in WMNs. OPSR introduces the opportunistic packet throughout metric. The forwarder is opportunistically selected as next relay node only it achieves *r*-times higher throughout than the gain that already accumulated by this packet. Once a node is chosen as forwarder for a received packet, this packet will be buffered in a priority-based forwarding queue. The packet with highest opportunistic gain is served first by the scheduler. A cross-layer cluster-based opportunistic routing has been proposed in [23]. In OPRL, the algorithm uses the cluster-tree structure of IEEE 802.15.4 standard for routing. Each node can associate with several parent nodes by taking advantage of an adequate organization of super frames at MAC layer. The next hop relay node is opportunistically chosen among multiple parent nodes based on its ETX value towards to sink. More opportunistic routing algorithms can be found in [9, 12-15, 17].

However, in these algorithms, the transmission power is fixed, which can not adapt the dynamic of network, especially in wireless sensor network where the network topology varies frequently. Therefore in [11] and [16], the authors introduce the transmission power adjustment into the opportunistic routing algorithms. In [11], the authors propose the transmission power control based opportunistic routing (TCOR) to save energy by reducing the transmission power of nodes, and to maintain communication reliability by employing opportunistic forwarding paradigm and leveraging the broadcast nature of wireless transmission medium. However, in TCOR, the authors do not take network interference and PDRsc into account. The similar algorithm can also be found in [16]. In [16], Mao et al. propose EEOR, which is an energy-efficient opportunistic routing protocol for WSNs. In EEOR, the candidate forwarders selection and prioritization schemes are optimized to minimize energy consumption. The sender keeps increasing its transmission power up to a maximum threshold, thereby increasing the number of neighbors. The relay nodes are sorted according to their energetic costs; the one that can be reached with the minimum expected energy consumption is selected by sender. However, the EEOR has the same disadvantages with TCOR.

## III. NETWORK MODEL

In this paper, the nodes in network are deployed uniformly [21]. Each node in network can communicate with other nodes whose distances to this node are smaller than its transmission range. For instance, as shown in Fig. 1, node *s* and node *r* can communicate with each other when $\|sr\| \leq r_s$, where $\|sr\|$ is the Euclidean distance between node *s* and node *r*, $r_s$ is the transmission range of node *s*. The nodes in network can adjust their transmission power from 0 to $P_{max}$. The coverage area of node *s* is a circle which the centre is node *s* and the radius is $r_s$, denoted as $C(s, r_s)$. This can be found in Fig. 1.

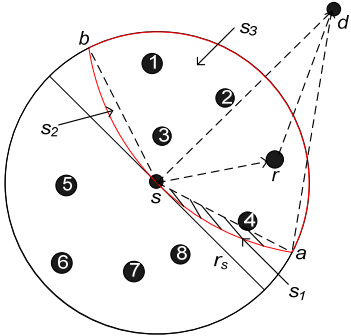

Fig. 1. The network model for opportunistic routing.

In opportunistic routing, when the source node wants transmit data packet to destination node, the source node relays the packet to a set of neighbor nodes rather than only one neighbor node. In this paper, we assume that the nodes in network know their locations and can exchange their locations periodically; the neighbor nodes whose distances to destination node are smaller than the source node have chance to be selected as relay nodes. This means that not all the neighbor nodes can be selected as relay nodes. Therefore, we define the candidate relay area as follows.

**Definition 1:** The candidate relay area of node *s* for transmitting data packet to destination node *d* is defined as the intersection area of two circles $C(s, r_s)$ and $C(d, \|ds\|)$, which is shown in Fig. 1 (the red area). $C(d, \|ds\|)$ means the circle which the center is the destination node *d* and the radius is $\|ds\|$.

Only the node in the candidate relay area can be selected as relay node. As shown in Fig. 1, the number of nodes in the coverage area of node *s* is defined as the node degree of node *s*. However, since not all the neighbors can be selected as relay nodes, therefore, we define the relay node degree of node *s* in Definition 3.

**Definition 2:** The candidate set of node *s* when transmits data packet to destination node *d* is defined as the set of nodes which locate in the candidate relay region, denoted as $R_{s \to d}(s, \|ds\|)$, which can be expressed as: $R_{s \to d}(s, \|ds\|) = \{1, 2, ..., i \mid \|di\| \leq \|ds\|\}$.

As shown in Fig. 1, the neighbors of node *s* are node 1 to node 8 and node *r*; however, according to Definition 2, only node 1 to node 4 and node *r* are the relay nodes in candidate set.

**Definition 3:** The relay node degree of node *s* is defined as the number of neighbors whose distances to the destination node are smaller than node *s*, i.e. the number of nodes in candidate relay region (shown in Fig. 1), denoted by $n_{rel}$.

According to Definition 3, in Fig. 1, the node degree of node *s* is 9 while the relay node degree is 5.

## IV. OPTIMIZATION MODEL

As shown in Section 1, two parameters can affect the performance of PDRsc, which are relay node degree and PDRsn. So in this section, we will investigate these two parameters in detail. Moreover, since the [11] and [16] shows that the excepted energy cost between the sender and the candidate set in opportunistic routing also relates to PDRsc, therefore, in this section, we also investigate the excepted energy cost under accurate PDRsc.

### A. Packet delivery ratio between sender and candidate set (PDRsc)

In opportunistic routing, the source node relays the data packet to all the nodes in candidate set, and the relay nodes transmit the data packet based on their relaying priorities [10]. There are two different kinds of PDR: 1) the PDR between sender and one relay node in candidate set, which can be found in Fig. 2(a); 2) the PDR between sender and candidate set, which can be found in Fig. 2(b).

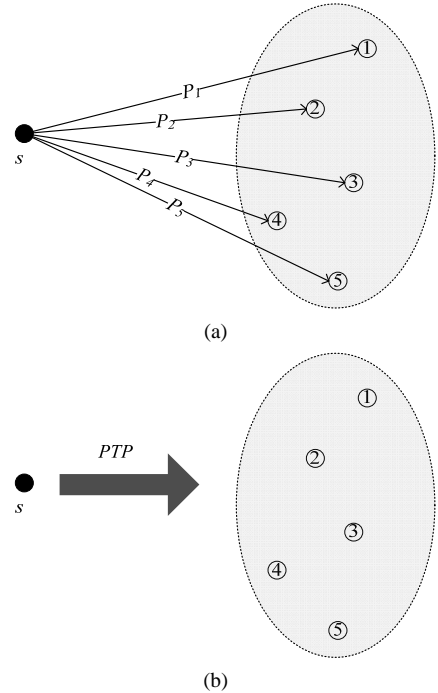

(a)

(b)

Fig. 2. Packet delivery ratio: (a) PDRsn; (b) PDRsc

Note that in opportunistic routing, the relay nodes in candidate set are all the one-hop neighbors of sender. Since the data packets are transmitted to more than one relay nodes in opportunistic routing, so the PDRsc increases comparing with the traditional routing algorithms. The PDRsc (as shown in Fig. 2(b)) can be calculated as:

$$PTP(P_{Ts}, n_{rel}) = 1 - \prod_{i=1}^{n_{rel}} (1 - p_i) \quad (1)$$

where $p_i$ is PDRsn and shown in Fig. 2(a); $n_{rel}$ is the relay node degree which is defined in Definition 3; $P_{Ts}$ is the transmission power of sender *s*.

According to (1), we can conclude that there are two parameters can affect the PDRsc: the relay node degree and the value of PDRsn. So in the following of this section, we will investigate the effection of these two parameters on the performance of PDRsc in detail.

## B. Packet delivery ratio between source node and relay node (PDRsn)

In wireless network, whether the packet can be received successfully by receiver relates to both the transmission power of sender and the interference of receiver [18]. The interference of receiver is defined as the summation of the interference nodes' transmission power (the interference node is defined as the node whose transmission ranges covers the receiver) [24]. This is the natural properties of wireless communication. The nodes not only can affect their neighbors, but also can be affected by interference nodes. Therefore, the $p_i$ shown in (1) is affected by both the transmission power and the interference.

In wireless network, if the receiver can decode the received data packet correctly in fading environment, one of the constraints is that the SINR (signal interference noise ratio) should above a certain threshold level $\beta$. According to [25], the probability that the SINR is above the given threshold can be calculated as:

$$P_{SINR}(P_{TS}) = P\left[\frac{\alpha_0^2}{\frac{1}{G}\sum_{i=1}^{m}\frac{K\alpha_i^2}{(\|ri\|/\|rs\|)^\eta}\frac{P_{Ti}}{P_{Ts}} + \frac{P_n\|rs\|^\eta}{P_{Ts}K}} \geq \beta\right]$$

$$= \int_0^\infty \int_0^\infty \cdots \int_0^\infty \exp\left(-\beta\left[\frac{1}{G}\sum_{i=1}^{m}\frac{K\alpha_i^2}{(\|ri\|/\|rs\|)^\eta}\frac{P_{Ti}}{P_{Ts}} + \frac{P_n\|rs\|^\eta}{P_{Ts}K}\right]\right)\prod_{i=1}^{m} e^{-\alpha_i}d\alpha_i$$

$$= \exp\left[-\frac{\beta P_n\|rs\|^\eta}{P_{Ts}K}\right]\prod_{i=1}^{m}\frac{1}{1+\frac{\beta P_{Ti}}{GP_{Ts}(\|ri\|/\|rs\|)^\eta}}$$

(2)

where $\alpha^2$ is an exponential random variable with unit mean; $\|ri\|$ is the distance between interference node and receiver; $\|rs\|$ is the distance between sender and receiver; $\eta$ is the propagation loss coefficient, where $2 \leq \eta \leq 5$; $K$ is the overall antenna gain and equals to $(G_tG_r\lambda^2)/((4\pi)^2 L)$; $G_t$ and $G_r$ are the transmission and reception antenna gain; $L$ is the system loss; $\lambda$ is the signal wavelength; $G$ is the processing gain; $P_n$ is the noise power at receiver, $P_{Ti}$ is the transmission power of *ith* interference node (except the sender *s*).

Additionally, if the data packet sent by sender can be received by receiver successfully, the transmission power of sender at receiver should larger than the receiving threshold. According to the conclusion in [24] and [25], the transmission power of sender at receiver can be calculated as:

$$P_R(P_{Ts}) = \left((K\alpha^2)/\|ds\|^\eta\right)P_{Ts} + P_n \quad (3)$$

where $P_n$ is the AWGN (additive white Gaussian noise) with zero mean and the variation is $\sigma^2$; $\|ds\|$ is the distance between sender *s* and receiver *r*.

Assuming that the receiving threshold which can guarantee correct data decoding at receiver is $P_{Thresh}$, then the probability that the transmission power of sender received by the receiver is equal to or larger than $P_{Thresh}$ can be calculated as [11]:

$$P = Q\left(\frac{1}{\sigma}\cdot\frac{P_{Thresh}\|ds\|^\eta}{P_{Ts}K\alpha^2}\right) \quad (4)$$

where $Q(x) = \frac{1}{\sqrt{2\pi}}\int_x^\infty e^{-(y^2/2)}dy$.

So according to (2) and (4), the PDRsn which takes both the transmission power loss and interference into account can be calculated, which is:

$$p_i = P_{SINR}(SINR \geq \beta)\cdot P_R(P_{Ts})$$

$$= Q\left(\frac{1}{\sigma}\cdot\frac{P_{Thresh}\|ds\|^\eta}{P_{Ts}K\alpha^2}\right)\cdot\exp\left[-\frac{\beta P_n\|rs\|^\eta}{P_{Ts}K}\right]\prod_{i=1}^{m}\frac{1}{1+\frac{\beta P_{Ti}}{GP_{Ts}(\|ri\|/\|rs\|)^\eta}} \quad (5)$$

## C. Relay node degree

As shown in (1), one parameter which can affect PDRsc is the relay node degree. The relay node degree has been defined in Definition 3. Moreover, the number of neighbors and the transmission power are relevant [21]. When the nodes are uniformly distributed in the event area, the probability that the number of neighbors is *n* for node *s* can be expressed as [21]:

$$P(n) = \left(((\rho\Psi)^n)/n!\right)e^{-\rho\Psi} \quad (6)$$

where $\Psi$ is the coverage area of node *s* and can be calculated by $\Psi = \pi r_s^2$, $r_s$ is the transmission range of node *s*; $\rho$ is the node density.

However, in ERTO, not the whole coverage area of sender is taken into account during the relay node selection; the interesting area is the candidate relay area which has been defined in Definition 1. So the (6) cannot be used directly to calculate the probability of relay node degree under specific transmission power.

The candidate relay area is shown in Fig. 1. In Fig. 1, the transmission range of sender *s* is $r_s$ and the distance between sender *s* and destination *d* is $\|ds\|$. The candidate relay area is the intersection area of two circles: $C(s, r_s)$ and $C(d, \|ds\|)$. So the length of line *da* shown in Fig. 1 equals to $\|ds\|$. The triangle *sda* is an isosceles triangle. Therefore, the angle of $\theta_{dsa}$ can be calculated as:

$$\theta_{dsa} = \arccos\left(r_s/(2\|ds\|)\right) \quad (7)$$

And the candidate relay area $\Delta$ can be calculated by:

$$\Delta = s_1 + s_2 + s_3 \quad (8)$$

where $s_3$ is the area of sector *sba* shown in Fig. 1; $s_1$ is the area of the shadow area; $s_2$ equals to $s_1$. Since the angle of $\theta_{dsa}$ has been calculated in (7), so the area of sector *sba* can be calculated as:

$$s_3 = r_s^2\cdot\arccos\left(r_s/(2\|ds\|)\right) \quad (9)$$

The value of $s_1$ equals to the area of sector *dsa* minus the area of triangle *dsa*, which can be expressed as:

$$s_1 = \frac{1}{2}\|ds\|^2\left[\left(\pi - 2\arccos\frac{r_s}{2\|ds\|}\right) - \sin\left(\pi - 2\arccos\frac{r_s}{2\|ds\|}\right)\right] \quad (10)$$

Since $s_1 = s_2$, so the candidate relay area $\Delta$ can be calculated as:

$$\Delta = r_s^2\cdot\arccos\frac{r_s}{2\|ds\|} + \|ds\|^2\left[\left(\pi - 2\arccos\frac{r_s}{2\|ds\|}\right) - \sin\left(\pi - 2\arccos\frac{r_s}{2\|ds\|}\right)\right]$$

(11)

Additionally, since the transmission range relates to the transmission power, so according to (3) and the transmission power of sender $s$, the transmission range $r_s$ can be expressed as:

$$r_s = \left(K\alpha^2 (P_{Thresh} - P_n) / P_{Ts}\right)^{1/\eta} \quad (12)$$

where $P_{Thresh}$ is the reception power threshold for successful data packet reception. Thus, the probability that there are $n_{rel}$ nodes in candidate relay area (i.e. the relay node degree is $n_{rel}$) can be calculated as:

$$P_{rnd}(P_{Ts}, n_{rel}) = \left(((\rho\Delta)^{n_{rel}}) / n_{rel}!\right) e^{-\rho\Delta} \quad (13)$$

where $\Delta$ and $r_s$ can be calculated based on (11) and (12). $P_{rnd}(P_{Ts}, n_{rel})$ means the probability that the relay node degree is $n_{rel}$ when the transmission power is $P_{Ts}$. Therefore, the selected relay node degree and transmission power of sender should make $P_{rnd}(P_{Ts}, n_{rel})$ as large as possible.

### D. Expected energy consumption

In this section, we introduce the expect energy consumption of the communication link between sender and candidate set. The excepted energy cost function between sender and candidate set has been proposed in [11] and [16]; however, the functions introduced in [11] and [16] are not accurate, since the authors do not take interference into account. Therefore, in this section, we investigate the expect energy consumption between sender and candidate set based on the conclusions in Section IV.A and Section IV.B.

Considering sender $s$ and its candidate set $R_{P_{Ts}}(s)$ with transmission power $P_{Ts}$. Let $\mathbb{C}_s(P_{Ts}, n_{rel})$ denotes the one-hop expected energy cost incurred by node $s$ with transmission power $P_{Ts}$ to send data packet which can be received by at least one node in candidate set $R_{P_{Ts}}(s)$. Therefore, according to [11] and [16], $\mathbb{C}_s(P_{Ts}, n_{rel})$ can be calculated as:

$$\mathbb{C}_s(P_{Ts}, n_{rel}) = E_s / \left(P_{sdr}(P_{Ts}, n_{rel})\right) \quad (14)$$

where $E_s$ is the energy that needed by transmitting and receipting the data packet which is transmitted from source node to candidate set. Based on the conclusion in [11] and [16], the $E_s$ can be calculated as:

$$E_s = \left(|R_{P_{Ts}}(s)| E_r + E_{P_{Ts}}\right) E\{T_{P_{Ts}}(s)\} \cdot (L/B) \quad (15)$$

where $E_r$ is the energy consumption for reception; $E_{P_{Ts}}$ is the energy consumption for transmitting at transmission power $P_{Ts}$; $L$ is the data packet size and $B$ is the bandwidth. In this paper, for simplifying the calculation, we assume that the $L$ and $B$ keep constant during the calculation. $E\{T_{P_{Ts}}(s)\}$ is the mean energy consumption that the packet transmitted by sender can be received by receiver successfully at $T_{P_{Ts}}(s)$ attempts, which can be calculated as [11]:

$$E\{T_{P_{Ts}}(s)\} = \sum_{l=0}^{\infty} l \times P_{tr}\{T_{P_{Ts}}(s) = l\} = 1 / \left(1 - \prod_{i=1}^{n_{rel}}(1 - p_i)\right) \quad (16)$$

where $P_{tr}\{T_{P_{Ts}}(s) = l\}$ is the probability that $T_{P_{Ts}}(s)$ equals to $l$ and can be calculated as: $P_{tr}\{T_{P_{Ts}}(s) = l\} = \prod_{i=1}^{n_{rel}}(1 - p_i)(1 - \prod_{i=1}^{n_{rel}}(1 - p_i))$.

Since $P_{sdr}(P_{Ts}, n_{rel})$ has been calculated by (5), so according to (15) and (16), the (14) can be rewritten as:

$$\mathbb{C}_s(P_{Ts}, n_{rel}) = \left((n_{rel} E_r + \xi P_{Ts})\delta\right) / \left(1 - \prod_{i=1}^{n_{rel}}(1 - p_i)\right)^2 \quad (17)$$

where $\delta = L/B$; $\xi$ is the consume coefficient for data packet transmission; $p_i$ can be calculated based on (5) which has taken transmission power loss and network interference into account.

## V. TOPOLOGY CONTROL BASED OPPORTUNISTIC ROUTING

### A. Optimal solution calculation

As shown in Section IV.A, in opportunistic routing, the PDRsc relates to both the relay node degree and the PDRsn. Since the larger relay node degree, the higher PDRsc is, so the relay node degree should as large as possible. However, large relay node degree needs large transmission power, which consumes more energy than the small transmission power. Unfortunately, the node energy are limited in wireless sensor networks and large transmission power cause more serious interference than the small one, so the transmission power should be minimized as far as possible. These are two opposite optimal objectives, and finding the optimal solutions for these issues have been proved are NP-hard problems [18]. This means that it is impossible to find an optimal solution which can make the relay node degree and PDRsc maximal while the network interference and energy consumption minimal at the same time. Therefore, in this paper, we introduce the multi-objective optimization into the optimal solutions calculation to find the tradeoff between these optimal objectives.

The issues introduced in Section IV can be expressed as:

$$\begin{cases} \max\left(PDR_{sc}(P_{Ts}, n_{rel}), P_{rnd}(P_{Ts}, n_{rel})\right) \\ \min\left(\mathbb{C}_s(P_{Ts}, n_{rel})\right) \end{cases}$$
$$st. \ 0 \le P_{Ts} \le P_{\max} \quad (18)$$
$$0 \le n_{rel} \le n$$

According to the multi-objective optimization theory [26][27], the (18) can be rewritten as:

$$\min_{x \in X^0} \mathbf{f}(\mathbf{x}) = (f_1(\mathbf{x}), f_2(\mathbf{x}), \ldots, f_m(\mathbf{x})) \quad (19)$$

where $\mathbf{X}^0 = \{\mathbf{x} \in \mathbb{R}^n | g_i(\mathbf{x}) \ge 0, i = 1, 2, \ldots, p\}$ is the feasible region; $\mathbf{x} = (x_1, x_2, \ldots, x_n)$ are the decision variables; $g_i(\mathbf{x}) \ge 0$, $i = 1, 2, \ldots, p$ are the constraint functions. In this paper, according to (18), there have:

$$\begin{cases} \mathbf{x} = (P_{Ts}, n_{rel}), \\ x_1 = P_{Ts}, \\ x_2 = n_{rel}; \end{cases} \quad (20)$$

$$\begin{cases} f(\mathbf{x}) = (-PDR_{sc}(P_{Ts}, n_{rel}), -P_{rnd}(P_{Ts}, n_{rel}), \mathbb{C}_s(P_{Ts}, n_{rel})), \\ f_1(\mathbf{x}) = -PDR_{sc}(P_{Ts}, n_{rel}), \\ f_2(\mathbf{x}) = -P_{rnd}(P_{Ts}, n_{rel}), \\ f_3(\mathbf{x}) = \mathbb{C}_s(P_{Ts}, n_{rel}); \end{cases} \quad (21)$$

$$\begin{cases} g_1(\mathbf{x}) = P_{\max} - P_{Ts}, \\ g_2(\mathbf{x}) = n - n_{rel}; \end{cases} \quad (22)$$

The $f_i(\mathbf{x})$ ($i = 1, 2, 3$) in (21) can be got according to (1), (13), and (17), respectively.

Different with the single-objective optimization, the optimal solution is a set rather than a single value in multi-objective optimization [26][27]. This solution set is named Pareto optimal solution set. The optimal solutions in Pareto optimal solution do not mean that they can satisfy all the optimal objectives shown in (18) simultaneous; these optimal solutions can improve the performance of the optimal objectives on at least one aspect compared with the solutions which are not in Pareto optimal solution set [26][27]. For instance, assuming $(P_{Ts1}, n_{rel1})$ is the optimal solution in Pareto optimal solution set and $(P_{Ts2}, n_{rel2})$ is not the optimal solution; if $PDR_{sc}(P_{Ts1}, n_{rel1}) > PDR_{sc}(P_{Ts2}, n_{rel2})$, then $P_{rnd}(P_{Ts1}, n_{rel1})$ and $\mathbb{C}_s(P_{Ts1}, n_{rel1})$ are at least not worse than $P_{rnd}(P_{Ts2}, n_{rel2})$ and $\mathbb{C}_s(P_{Ts2}, n_{rel2})$, respectively. The optimal solutions in Pareto optimal solution set do not have comparability [26][28][29]. For instance, assuming that $(P_{Ts1}, n_{rel1})$ and $(P_{Ts2}, n_{rel2})$ are all the optimal solutions in Pareto optimal solution set; if $PDR_{sc}(P_{Ts1}, n_{rel1}) > PDR_{sc}(P_{Ts2}, n_{rel2})$, then at least one parameter of $P_{rnd}(P_{Ts1}, n_{rel1})$ and $\mathbb{C}_s(P_{Ts1}, n_{rel1})$ is worse than that of $P_{rnd}(P_{Ts2}, n_{rel2})$ and $\mathbb{C}_s(P_{Ts2}, n_{rel2})$.

According to [26] and [27], the definition of Pareto optimal solutions for this issue is defined as follows:

**Definition 4.** Assuming that $(\bar{P}_{Ts}, \bar{n}_{rel}) \in \mathbf{X^0}$, if there is no $(P_{Ts}, n_{rel}) \in \mathbf{X^0}$ which can make $f_i((P_{Ts}, n_{rel})) < f_i((\bar{P}_{Ts}, \bar{n}_{rel}))$ ($i = 1, 2, 3, 4$) hold, then $(\bar{P}_{Ts}, \bar{n}_{rel})$ is the Pareto Optimal Solution of (18), the set of all Pareto Optimal Solutions is Pareto Optimal Solution Set, denoted as $\mathbf{R}$.

There are two different kinds of algorithms to calculate the Pareto optimal solution set of multi-objective optimization [26]: 1) the traditional algorithm; for instance, method of objective weighting, method of distance functions, min-max formulation, etc.; the drawbacks of traditional algorithms have been introduced in [26]; 2) the intelligent optimization algorithm; the intelligent optimization algorithm includes the evolutionary algorithm, particle swarm optimization, etc. Since the calculation of Pareto optimal solution set is not the main research item of this paper, so we use the evolutionary algorithm introduced in [27], which is more accuracy and efficiency than traditional algorithm, to calculate the Pareto optimal solution set of this issues. The detail of this algorithm can be found in [27].

According to Definition 4, there is more than one solution in Pareto optimal solution set, so the Pareto optimal solution set $\mathbf{R}$ can be expressed as:

$$\mathbf{R} = \left\{ (\bar{P}_{Ts1}, \bar{n}_{rel1}), (\bar{P}_{Ts2}, \bar{n}_{rel2}), \cdots, (\bar{P}_{Tsn}, \bar{n}_{reln}) \right\} \quad (23)$$

The solutions shown in (23) can be chosen as the optimal solution of (18).

*B. Topology control algorithm*

Once the routing process begins, first, each node executes a fully distributed algorithm to collect the required information, which can be utilized to estimate PDRsc and expected energy consumption. The information which is needed can be collected through iterative one-hop beacons. The local information is updated by the latest time-stamp in these collection methods. When nodes get the required information, they will calculate the Pareto optimal solution set for topology control. Based on the algorithm introduced in [27], the Pareto optimal solution set $\mathbf{R}$ of issue (18) can be gotten. According to the characteristic of the solutions in Pareto optimal solution set, which has been introduced in Section V.A, any solution in Pareto optimal solution set can be chosen as the final optimal solution. However, in practice, each node only has one transmission power and relay node degree, which means that not all the solutions in $\mathbf{R}$ can be selected. Therefore, according to the Pareto optimal solution set and the current transmission power and relay node degree, there are two different topology control strategies as follows.

1. The current transmission power and relay nod degree are not in Pareto optimal solution set $\mathbf{R}$.

In this situation, since the current transmission power and relay nod degree are not in Pareto optimal solution set $\mathbf{R}$, so the transmission power needs to be adjusted.

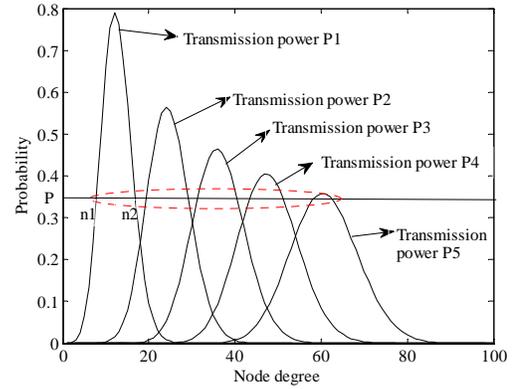

Fig. 3. Relationship between node degree and transmission power

As shown in [17], the relationship between transmission power and relay node degree is probabilistic in physical layer, i.e. when the transmission power is $P_{Ts}^*$, the relay node degree is $n_{rel}^*$ with probability $P_{rnd}(P_{Ts}^*, n_{rel}^*)$, which is shown in (13) and Fig. 3. This means that the optimal solutions $(P_{Ts}, n_{rel})$ in $\mathbf{R}$ may not exist or exist with low probability in reality. This can be found in Fig. 3. For instance, when the current relay node degree is $n_{rel1}$, if the relay node degree need to be reduced to $n_{rel2}$, then the needed transmission power $P_{Ts}$ is probabilistic and not fixed. Moreover, as shown in Fig. 3, assuming that $(P_{Ts1}, n_{rel1}) \in \mathbf{R}$ can make the PDRsc maximal and the expected energy consumption minimal at the same time, but the value of the probability shown in (13) is very low, then if $P_{Ts1}$ is chosen as optimal transmission power, the relay node degree may not $n_{rel1}$ with high probability; so the performance of PDRsc and expected energy consumption will deteriorate. Therefore, we choose the optimal solutions in $\mathbf{R}$ which can make (13) get the maximum value as the fianl optimal solution of (18).

Another issue needs to be solved during the optimal transmission power and relay node degree selection is that the non-uniqueness of (13), which can be found in Fig. 3. As shown in Fig. 3, if the maximum probability shown in (13) that calculated based on the solutions in Pareto optimal solution set is $P$, then the solutions $(P_{Ts}, n_{rel})$ are not uniqueness, such as the solutions in the red area of Fig. 3 have the same probability $P$. So the optimal transmission power needed to be decided based on the values of PDRsc and expected energy consumption during these solutions. Therefore, we define the optimal feasible solution set as follows.

**Definition 5.** When the maximum probability shown in (13) is $P$ which is calculated based on the solutions in Pareto optimal solution set, then the solutions which can make the probability shown in (13) equal to $P$ are the elements of optimal feasible solution set, noted as $R^*$, where $R^* \in \mathbf{R}$.

For instance, as shown in Fig.3, the solutions in the red area are all the elements of optimal feasible solution set.

According to Definition 5, the feasible region has been reduced from Pareto optimal solution set to optimal feasible solution set. The optimal transmission power and relay node degree will be chosen from the optimal feasible solution set $R^*$. Considering the fact that the node performance will be decided by the worst node parameter, which is called the cask theory, so for getting a balanced solution, we propose the balanced optimal solution selection algorithm as follows.

In balanced optimal solution selection algorithm, when the optimal solution is $(P_{Tsi}, n_{reli})$, then the corresponding PDRsc, the probability shown in (13), and the expected energy consumption will be $PDR_{sci}(P_{Tsi}, n_{reli})$, $P_{rndi}(P_{Tsi}, n_{reli})$, $\mathbb{C}_{si}(P_{Tsi}, n_{reli})$, respectively. For different optimal solutions, these values are different. The optimal feasible solution set is $[(P_{Ts1}, n_{rel1}), (P_{Ts2}, n_{rel2}), \cdots, (P_{Tsm}, n_{relm})]$, and the corresponding performance matrix can be expressed as (since the probabilities shown in (13) are equal in optimal feasible solution set, so this matrix does not include it): $[(PDR_{sc1}, \mathbb{C}_{s1}), (PDR_{sc2}, \mathbb{C}_{s2}), \cdots, (PDR_{scm}, \mathbb{C}_{sm})]$, where $m$ is the number of solutions in optimal feasible solution set. The variance matrix of PDRsc and expected energy consumption can be expressed as: $v = [v_{PDRsc}, v_{\mathbb{C}_s}]$. The parameters (i.e. PDRsc and expected energy consumption) which the variance is larger will have greater effect on the optimal solution selection than that of the smaller one.

Since the optimal solutions in Pareto solution set do not have comparability and the probabilities are equal in optima feasible solution set, so the PDRsc and expected energy consumption shown in the performance matrix have properties as follows.

**Corollary 1.** In optimal feasible solution set, if $PDR_{sci}(P_{Tsi}, n_{reli}) > PDR_{scj}(P_{Tsj}, n_{relj})$ or $PDR_{sci}(P_{Tsi}, n_{reli}) < PDR_{scj}(P_{Tsj}, n_{relj})$, then there must exist $\mathbb{C}_{si}(P_{Tsi}, n_{reli}) > \mathbb{C}_{sj}(P_{Tsj}, n_{relj})$ or $\mathbb{C}_{si}(P_{Tsi}, n_{reli}) < \mathbb{C}_{sj}(P_{Tsj}, n_{relj})$, $i, j \in (1, 2, 3, ..., m)$, respectively; vice versa. Moreover, if $PDR_{sci}(P_{Tsi}, n_{reli}) = PDR_{scj}(P_{Tsj}, n_{relj})$, then there must exist $\mathbb{C}_{si}(P_{Tsi}, n_{reli}) \neq \mathbb{C}_{sj}(P_{Tsj}, n_{relj})$, vice versa.

Proof. As shown in (18), the purpose of multi-objective optimization is to get large PDRsc and small expected energy consumption as far as possible. Assuming that $(P_{Ts1}, n_{rel1})$ and $(P_{Ts2}, n_{rel2})$ are in optimal feasible solution set $R^*$, then $P_{rnd1}(P_{Ts1}, n_{rel1})$ equals to $P_{rnd2}(P_{Ts2}, n_{rel2})$. Moreover, since $R^* \in \mathbf{R}$, so according to Definition 5, to $PDR_{sc1}(P_{Ts1}, n_{rel1})$, $\mathbb{C}_{s1}(P_{Ts1}, n_{rel1})$, and $P_{rnd1}(P_{Ts1}, n_{rel1})$, at least one of these parameters is better than that of $PDR_{sc2}(P_{Ts2}, n_{rel2})$, $\mathbb{C}_{s2}(P_{Ts2}, n_{rel2})$, and $P_{rnd2}(P_{Ts2}, n_{rel2})$; meanwhile, at least one parameter of $PDR_{sc1}(P_{Ts1}, n_{rel1})$, $\mathbb{C}_{s1}(P_{Ts1}, n_{rel1})$, and $P_{rnd1}(P_{Ts1}, n_{rel1})$ is worse than that of $PDR_{sc2}(P_{Ts2}, n_{rel2})$, $\mathbb{C}_{s2}(P_{Ts2}, n_{rel2})$, and $P_{rnd2}(P_{Ts2}, n_{rel2})$. Since $P_{rnd1}(P_{Ts1}, n_{rel1})$ is equal to $P_{rnd2}(P_{Ts2}, n_{rel2})$, so the conclusion in Corollary 1 holds.

According to the conclusion of Corollary 1, we can conclude the Corollary 2 as follows.

**Corollary 2.** In optimal feasible solution set, the order of PDRsc from large to small is the same as that of the corresponding expected energy consumption, i.e. large PDRsc means large expected energy consumption.

Proof. The meaning of Corollary 2 can be explained as follows: in $R^*$, when PDRsc is the largest, then the corresponding expected energy consumption is the largest, too; when PDRsc is the second largest, then the corresponding expected energy consumption is the second largest; and so on. For proofing Corollary 2, we assume that the second largest PDRsc is $PDR_{sc2}(P_{Ts2}, n_{rel2})$ and the corresponding expected energy consumption is $\mathbb{C}_{s2}(P_{Ts2}, n_{rel2})$. If $\mathbb{C}_{s2}(P_{Ts2}, n_{rel2})$ is not the second largest, then according to the properties of Pareto optimal solutions [26], there must exist $PDR_{scx}(P_{Tsx}, n_{relx})$ and $\mathbb{C}_{sx}(P_{Tsx}, n_{relx})$ which satisfy $PDR_{sc2}(P_{Ts2}, n_{rel2}) > PDR_{scx}(P_{Tsx}, n_{relx})$ and $\mathbb{C}_{s2}(P_{Ts2}, n_{rel2}) < \mathbb{C}_{sx}(P_{Tsx}, n_{relx})$; this conclusion does not conform the conclusion of Corollary 1. So the Corollary 2 is proved.

The conclusion of Corollary 2 means that in $R^*$, when PDRsc is large, then the corresponding expected energy consumption is large, too; vice versa. However, our purpose is to find an optimal solution that can make the PDRsc is the largest while the expected energy consumption is the smallest, which is impossible in $R^*$ according to the conclusion of Corollary 2. So we need find a tradeoff between PDRsc and expected energy consumption.

For getting the balanced optimal solution, based on the conclusion of Corollary 2, we choose the intermediate value of PDRsc and corresponding expected energy consumption in optimal feasible solution set as the optimal solution, since these solutions are more balanced than the other solutions in optimal feasible solution set. The intermediate value can be calculated as:
1. when the number of solutions in $R^*$ is odd, then the optimal solution is $(P_{Ts(m+1)/2}, n_{rel(m+1)/2})$, where $m$ is the number of solutions in $R^*$;
2. when $m$ is even, two optimal solutions can be gotten: $(PDR_{scm/2}(P_{Ts(m/2)}, n_{rel(m/2)}), \mathbb{C}_{s(m/2)}(P_{Ts(m/2)}, n_{rel(m/2)}))$ and $(PDR_{sc(m+2)/2}(P_{Ts(m+2)/2}, n_{rel(m+2)/2}), \mathbb{C}_{s(m+2)/2}(P_{Ts(m+2)/2}, n_{rel(m+2)/2}))$.
According to the fact that the parameter which the variance is larger has greater effect on the optimal solution selection than the smaller parameter, so the parameter which its variance and value are large will be chosen as the optimal solution. For instance, if $v_{PDRsc} > v_{\mathbb{C}_s}$ and $PDR_{scm/2}(P_{Ts(m/2)}, n_{rel(m/2)}) > PDR_{sc(m+2)/2}(P_{Ts(m+2)/2}, n_{rel(m+2)/2})$, then $(P_{Ts(m/2)}, n_{rel(m/2)})$ will be chosen as the optimal solution.

The balanced optimal solution selection algorithm can be found in Algorithm 1.

---

**Algorithm 1.** Balanced optimal solution selection algorithm

1. Calculating the optimal feasible solution set $R^*$;
2. Calculating the $v_{PTP}$ and $v_{\mathbb{C}_s}$ of PDRsc and expected energy

consumption in optimal feasible solution set $R^*$, respectively;
3. if $m$ is odd
4.     $P \leftarrow P_{Ts(m+1)/2}$ and $n_{rel} \leftarrow n_{rel(m+1)/2}$;
5. else if $m$ is even
6.     if $v_{PTP} > v_{\mathbb{C}_s}$ and $PDR_{scm/2}(P_{Ts(m/2)}, n_{rel(m/2)})$ larger than $PDR_{sc(m+2)/2}(P_{Ts(m+2)/2}, n_{rel(m+2)/2})$
7.         $P \leftarrow P_{Ts(m/2)}$ and $n_{rel} \leftarrow n_{rel(m/2)}$;
8.     else
9.         $P \leftarrow P_{Ts(m+2)/2}$ and $n_{rel} \leftarrow n_{rel(m+2)/2}$;
10.     end if
11.     if $v_{PTP} < v_{\mathbb{C}_s}$ and $\mathbb{C}_{s(m/2)}(P_{Ts(m/2)}, n_{rel(m/2)})$ larger than $\mathbb{C}_{s(m+2)/2}(P_{Ts(m+2)/2}, n_{rel(m+2)/2})$
12.         $P \leftarrow P_{Ts(m+2)/2}$ and $n_{rel} \leftarrow n_{rel(m+2)/2}$;
13.     else
14.         $P \leftarrow P_{Ts(m/2)}$ and $n_{rel} \leftarrow n_{rel(m/2)}$;
15.     end if
16. end if

2. The current transmission power and relay node degree are in Pareto optimal solution set **R**.

In this situation, since the current transmission power and relay node degree are in Pareto optimal solution set **R**, which are better than the solutions that not in Pareto solution set, so for reducing the control cost, the nodes do not adjust their transmission power.

The reasons that the nodes in this scenario do not adjust their transmission powers are: on one hand, the network topology in wireless sensor network changes frequently due to node mobility or node failure, so the frequent transmission power adjustment will consume large amount of network resources that could have been used in packet transmission; on the other hand, the current transmission power and relay node degree in Parte optimal solution set **R** means that the current network performance is good. So considering the energy consumption and control cost by controlling the network topology, this tradeoff is worthy.

Based on the conclusions above, the process of transmission power adjustment can be shown as follows.

**Algorithm 2.** Transmission power adjustment algorithm
1. Source node collects the needed information from neighbor nodes;
2. Source node calculates the Pareto optimal solution set **R** according to these information;
3. if $(P_{Ts}, n_{rel}) \in \mathbf{R}$
4.     $P \leftarrow P_{Ts}$;
5.     $n \leftarrow n_{rel}$;
6. else if $(P_{Ts}, n_{rel}) \notin \mathbf{R}$
7.     call Algorithm 1;
8.     end if
9. end if

*C. Efficient and reliable topology control based opportunistic routing algorithm*

When the optimal transmission power and relay node degree have been gotten, the sender adjusts the transmission power and transmits data packet to the candidate set. For getting high PDRsc, all the nodes in candidate relay area will be chosen as relay nodes in candidate set. Because of the PDRsn of each relay node has been calculated during the topology control, which can be found in Fig. 2(a) and (5), so for reducing the computation complexity, in ERTO, we use except transmission account (ETX) as the performance matrix to calculate the priority of each relay node in candidate set. The ETX used in this paper is different with the traditional definition of ETX which can be found in ExOR, since in this paper, the calculation of ETX takes both the network interference and transmission power loss into account. Therefore, the ETX can be calculated as:

$$ETX(i) = 1 / p_i \qquad (24)$$

where $p_i$ can be calculated by (5).

In candidate set, the node which has small ETX has high priority for data packet transmission. Similar to that shown in ExOR, this process continuous until the data packet is received by destination node.

The efficient and reliable topology control based opportunistic routing algorithm is shown in Algorithm 3.

**Algorithm 3:** Efficient and reliable topology control based opportunistic routing algorithm
1. Source node calculate the Pareto optimal solution set for transmission power and relay node degree based on the algorithm introduced in (31);
2. Source node adjusts the transmission power according to Algorithm 2;
3. Source node calculate the ETX for each relay nodes in relay node set based on the probability $p_i$;
4. Source node calculate the priorities for each relay nodes in relay node set based on the value of ETX;
5. Source node transmits data packet to all the nodes in candidate relay area;
6. Relay nodes relay data packet to next hop relay nodes based on their priorities;
7. Repeating Step 1 to Step 6 until the data packet received by destination node.

VI. SIMULATION AND DISCUSSION

*A. Simulation Configuration*

In this section, the performance of efficient and reliable topology control based opportunistic routing algorithm is presented. For comparing the performance of ERTO, three opportunistic routing algorithms are implemented in this simulation: 1) ExOR [10]; 2) TCOR [11]; 3) EEOR [16]. ExOR is the traditional opportunistic routing algorithm without power control and the performance matrix is ETX; the TCOR and EEOR are power control based opportunistic routing algorithms and the performance matrix of these two algorithms is the expected energy consumption between sender and relay node set. The results are shown in Fig. 4-Fig. 10.

In simulation, the nodes are distributed in the event area uniformly. Each node knows their own location and can exchange their location periodically. Moreover, the antenna of node is the omnidirectional antenna in which the transmission range can be changed based on the transmission power. The constant bit rate (CBR) [19][30][31] is used in this paper to generate data packet; each CBR data packet is transmitted between two nodes which are chosen randomly. The number of CBR connection pairs represents the traffic load in network. The larger number of CBR connection pairs are, the higher traffic load is. More simulation parameters can be found in Table 1.

TABLE 1. SIMULATION CONFIGURATION

| Parameter | Value |
|---|---|
| deployment area | 1000*m*×1000*m* |
| initial node transmission range | 100*m* |
| packet length | 1024bits |
| data rate | 15Kbps |
| initial energy | 5J |
| maximum transmission power | 0.8W |
| minimum transmission power | 0.1W |
| number of nodes | 40-120 |
| receiving power | 0.05W |
| simulation time | 300*s* |
| number of CBR pairs | 20-100 |
| simulation tool | NS-2 |

The varying parameters during the simulation are the number of nodes in network, the number of CBR connection pairs, and the simulation time. The performance metrics used in this simulation are introduced briefly:

1. *Packet delivery ratio*. As defined in [32] and [33], the packet delivery ratio represents the ratio of all successfully received data packets at receiver to the total number of data packets generated by the application layer at source node.

2. *Transmission delay*. This is the transmission delay of data packet from source node to destination node; the end-to-end delay includes the queuing delay, the delay caused by re-transmission, and the packet-carrying delay [32][33].

3. *Network throughput*. As the definition in [34], the network throughput is the ratio of the total number of data packets successfully received by destination node to the number of data packets sent by all the nodes during the simulation time.

4. *Residual energy*. Different with the traditional definition of residual energy, the residual energy used in this paper is defined as the ratio of the residual energy of node to the total energy of node, which can be expressed as: $R_p = \frac{residual\ energy}{total\ energy}$.

### B. Performance under different number of nodes

In this section, the performance of ERTO under different node densities is presented, which can be found from Fig. 4 to Fig. 7. In this simulation, the number of CBR connection pairs is fixed and equals to 30.

Fig. 4 illustrates the packet delivery ratio of these four opportunistic routing algorithms. In Fig. 4, the packet delivery ratio of ERTO is much higher than that of the other three algorithms, such as 40% higher than ExOR and 20% higher than EEOR when the number of node is 100. With the increasing of node number, the packet delivery ratio increases obviously in ExOR and ERTO while the increasing is slight in TCOR and EEOR. Moreover, when the number of node is large, the increasing in ExOR is smaller than that in ERTO. Two parameters can be used to explain this conclusion: the node degree and the network interference. When the network density is small, the node degree is the domain parameter on determining the packet delivery ratio; when the node degree increases, the packet delivery ratio increases. However, with the increasing of the node density, the network interference becomes more and more serious than that in sparse network. So when the network density is large, the network interference will be the domain parameter. Since the ERTO takes the network interference into account, so even in density network, the packet delivery ratio of ERTO increases.

The performance of packet delivery ratio can also affect the performance of end-to-end delay, which has been presented in Fig. 5. On one hand, the higher packet delivery ratio means lower probability of packet retransmission, which reduces the transmission delay; so with the increasing of node density, the transmission delay decreases. On the other hand, the number of transmission hops can also affect the performance of end-to-end delay; the large hop numbers can increase the transmission delay. In ExOR, since the transmission power cannot be changed, so with the increasing of node density, the transmission hops increases slower than that in the other three algorithms. Therefore, considering both the packet delivery ratio and the number of transmission hops, the decreasing of transmission delay in ExOR is larger than that of the other three algorithms. Moreover, since the packet delivery ratio in ERTO is better than that in ExOR, EEOR, and TCOR, so the transmission delay in ERTO is the smallest.

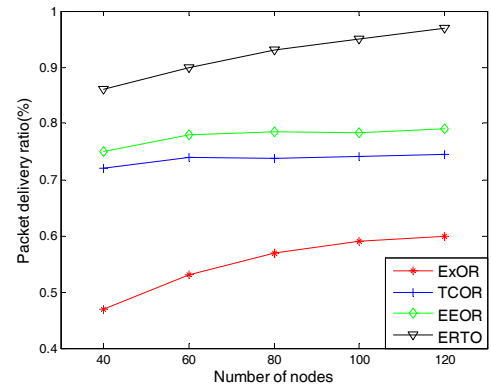

Fig. 4. Packet delivery ratio under different node densities

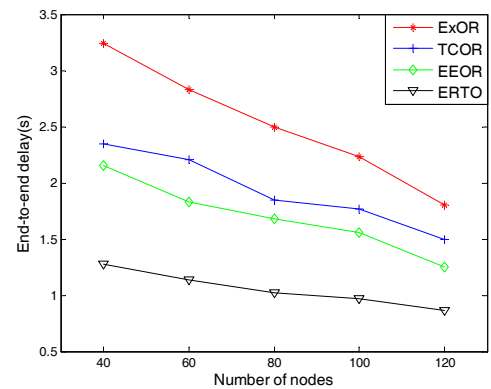

Fig. 5. End-to-end delay under different node densities

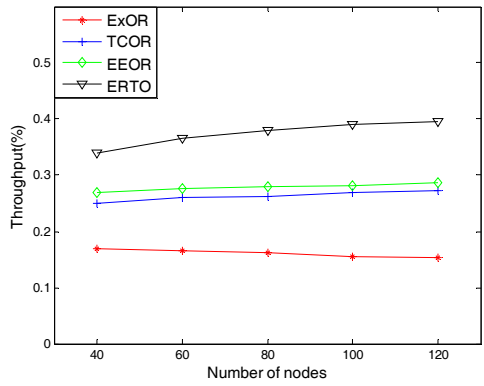

Fig. 6. Throughput under different node densities

The outstanding performance of packet delivery ratio and end-to-end delay in ERTO also contributes to improve the comprehensive performance of the network, such as the network throughput, which has been shown in Fig. 6.

In Fig. 6, the performance of network throughput has been presented. The throughput of ERTO is the highest among these four algorithms; and with the increasing of network density, the increasing of throughput is slight in these four algorithms. The throughput of ERTO increases larger with the increasing of node density than the other three algorithms. This can be explained as follows. On one hand, when the network density increases, the packet delivery ratio increases, which can be found in Fig. 4, so the throughput increases with the increasing of node density; on the other hand, the more nodes in network, the more serious interference which deteriorates the performance of throughput. Considering these two aspects, the throughput in ExOR, TCOR, and EEOR is stable; however, because ERTO has taken network interference into account, so the throughput increases obviously when the network density increases.

## B. Performance under different traffic load

In this section, the performance of these four algorithms under different traffic load is presented. In this simulation, we use the number of CBR connection pairs to represent the different traffic load [19][30][31]. The results can be found in Fig. 7, Fig. 8, and Fig. 9. In this simulation, the number of nodes in network is 100.

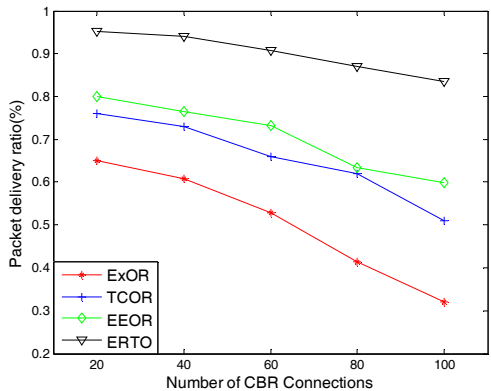

Fig. 7. Packet delivery ratio under different traffic load

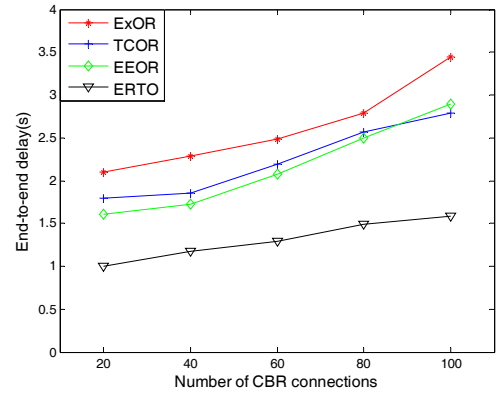

Fig. 8. End-to-end delay under different traffic load

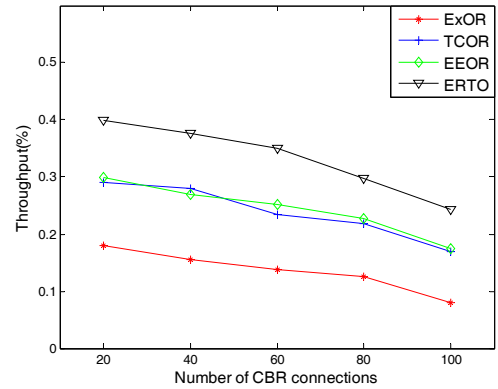

Fig. 9. Throughput under different traffic load

Actually, when the traffic load in network increases, the most important parameters which can affect the network performance greatly are the network congestion and contention. The more serious network congestion and contention, the worse network performance is, which can be found in Fig. 7 and Fig. 8. In Fig. 7, with the increasing of traffic load, both the packet delivery ratios of these four algorithms decrease; moreover, the decreasing ratio is the largest in ExOR while it is the smallest in ERTO. Similarly results can also be found in Fig. 8. In Fig. 8, when the traffic load increases, on one hand, the network congestion and contention increase, so the transmission delay increases; on the other hand, due to the packet delivery ratio decreases, the transmission delay increases further. However, since ERTO can adjust the transmission power during the routing process base on network interference, so the performance of packet delivery ratio and transmission delay are all better than that of the other three algorithms.

The throughput of these four algorithms under different traffic load has been shown in Fig. 9. In Fig. 9, when the traffic load increases, the throughput decreases. These can be explained as: 1) similar to the performance of packet delivery ratio and transmission delay, when the traffic load increases, the network congestion and contention increases, which deteriorates the performance of network throughput; 2) when the traffic load increases, both the packet delivery ratio and transmission delay become worse than that when the traffic load is small, so the throughput decreases further. However, since the performance of packet delivery ratio and the transmission delay in ERTO are the best, and ERTO takes the network interference into account during the topology control, so the performance of throughput in ERTO is also the best.

*C. Performance of energy consumption under different simulation time*

The energy consumption of ERTO is better than that of the other three algorithms, which is present in Fig. 10. The number of CBR connection pairs is 30 and the number of nodes is 100 in this simulation. In Fig. 10, with the increasing of simulation time, both the residual energy of these four algorithms reduces. However, the reduction of ERTO is smaller than that of ExOR, EEOR, and TCOR. Since the ExOR can not adjust the transmission power and does not take energy consumption into account, so the reduction of ExOR is fastest in these four algorithms. Even EEOR, TCOR, and ERTO all take energy consumption into account during the routing process, considering the performance of packet delivery ratio and transmission delay, the energy consumption in ERTO is better than that in EEOR and TCOR.

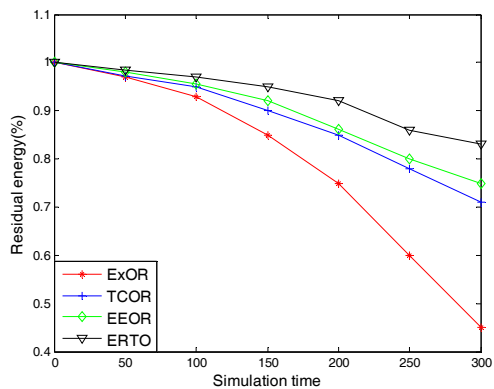

Fig. 10. Residual energy of ExOR, TCOR, EEOR, and ERTO

VII. CONCLUSION

In this paper, we propose an efficient and reliable topology control based opportunistic routing algorithm for wireless sensor network. Different with the previous works, in ERTO, the packet delivery ratio between source node and candidate set, the energy consumption, and the relay node degree have been taken into account to find the optimal transmission power and relay node degree. Considering the fact that finding the optimal solutions which can meet the requirements of all the issues introduced in Section IV is NP-hard problem, in this paper, we introduce the multi-objective optimization into the algorithm. The optimal transmission power and the relay node degree will be decided based on the properties of the optimal solutions in Pareto optimal solution set. Based on the innovations above, the performance of opportunistic routing has been improved greatly. The energy consumption, the transmission delay, the throughout, and the packet delivery ratio have been improved remarkably compared with ExOR, TCOR, and EEOR.